# Vortex solitons in large-scale waveguide arrays with adjustable discrete rotational symmetry

Yaroslav V. Kartashov

*Institute of Spectroscopy, Russian Academy of Sciences, 108840, Troitsk, Moscow, Russia*
*kartashov@isan.troitsk.ru*

We consider vortex solitons in large-scale arrays composed of $N$ elliptical waveguides placed on a ring, which can be fabricated using fs-laser writing technique in transparent nonlinear dielectrics. By introducing variable twist angles between longer axes of neighboring elliptical waveguides on a ring, we create circular arrays with adjustable discrete rotational symmetry ranging from $\mathcal{C}_N$ to $\mathcal{C}_1$, when the number of waveguides $N$ on the ring remains fixed. This allows to consider the impact of discrete rotational symmetry on the properties of available vortex solitons without changing the number of guiding channels in the structure, and to predict how exactly splitting of higher-order phase singularities into sets of charge-1 singularities occurs in vortex states, when they are forbidden by the discrete rotational symmetry of the structure that imposes the restrictions on the maximal possible vortex charge. It is found that separation between split charge-1 phase singularities in such higher-order vortex states increases with increase of the order of solution. We also study linear spectra of such arrays and show how variation of their discrete rotational symmetry affects linear eigenmodes, whose combinations can give rise to vortex modes. We also show that variation of discrete rotational symmetry in arrays with fixed number of guiding channels $N$ has strong impact on stability of vortex solitons. Thus, only higher-charge vortex solitons are stable in such large-scale arrays and the number of stable states typically decreases with decrease of the order of discrete rotational symmetry of the structure at fixed $N$.

**Keywords:** Vortex solitons; waveguide arrays; discrete rotational symmetry.

The properties of self-sustained nonlinear excitations (solitons) that propagate in the nonlinear optical medium without changes in their shape strongly depend on the spatial symmetry of the underlying optical material. This is the case both for the simplest fundamental solitons and for excited states, such as multipole and vortex solitons [1]. For instance, in uniform optical materials Galilean invariance allows us to construct states moving with different transverse velocities, while the azimuthal symmetry allows, in principle, the formation of vortex-ring solitons with any topological charge. Transverse refractive index modulation, even if it is shallow, dramatically modifies light propagation dynamics in both linear and nonlinear regimes [2,3], because it may break translational or azimuthal symmetries of the system. Particularly interesting situation is realized when refractive index landscape is periodic in the transverse direction [4-7], so that Bloch waves representing eigenstates of periodic medium replace plane waves – eigenmodes of the uniform medium. Accordingly, the dynamics of linear wavepackets in optical lattices are determined by Bloch waves that they excite, while bandgap structure of the spectrum resulting from periodic refractive index now dictates the domains, where solitons can form in the presence of nonlinearity. One of the most remarkable advantages of periodic optical lattices is that they have strong stabilizing action on self-sustained states allowing to observe them in stable form in multidimensional geometries [8-17], not only in the form of fundamental states, but also in the form of soliton complexes. For instance, optical lattices suppress azimuthal modulation instabilities that usually destroy ring-like vortex solitons in uniform medium, leading to the formation of stable vortex lattice solitons predicted in discrete [18] and continuous [13] lattices, and subsequently experimentally observed [19-23] and theoretically studied [24-27] in lattices with different symmetries, see recent reviews [28-30].

Azimuthally periodic lattices characterized by discrete rotational symmetry that can be optically induced using modulated variants or the interference patterns of Bessel beams [31-35] or created technologically as photonic crystal fibers [36-39] also support rich varieties of vortex solitons. Their spectrum can be analyzed using phenomenology of the angular Bloch modes that represent angularly periodic eigenstates of such systems [40,41]. Thus, symmetry analysis based on the properties of the angular Bloch modes of azimuthal lattices has revealed fundamental restrictions imposed by discrete rotational symmetry of the lattice on the available charges of vortex solitons that can exist in such structures [42,43], at least when their intensity distributions feature the same order of the discrete rotational symmetry as that of the lattice. It is due to these restrictions, the experiments with square lattices reveal the formation of only charge-1 vortex solitons [19,20], while in experiments with triangular lattices it was possible to observe solitons with topological charges up to 2 [23]. In addition, the discrete rotational symmetry of the medium determines stability properties of vortex solitons, leading to unexpected stability of states with highest charges and strong instabilities of low-charge solitons. Stability properties of vortex solitons in lattices with different discrete rotational symmetries can be strongly affected by dissipative effects [44-46] and dynamical rotations of the underlying lattice [47]. Analysis of the properties of nonlinear excitations in azimuthally periodic guiding structures with different discrete rotational symmetries shows that in addition to vortices [48-51] these structures support various types of gap, multipole, and circulating states [52-54], they can be used for realization of angular switching [55,56], efficient pulse combining [57], and realization of unusual angular diffraction scenarios [58,59].

At the same time, the interplay between discrete rotational symmetry and the properties of vortex solitons has so far been considered only in structures consisting of identical waveguides, where the order of discrete rotational symmetry of the structure exactly coincides with the number of waveguides forming the structure. But what happens if the system – for example, circular waveguide array – consists of large number of guiding elements, while its discrete rotational symmetry is substantially lower than the number of guiding elements? To show how the properties of vortex solitons change in this case, we consider large-scale waveguide arrays consisting of $N$ elliptical waveguides placed on a ring, where discrete rotational symmetry can be controllably reduced from $\mathcal{C}_N$ to $\mathcal{C}_1$ by varying mutual twist angle between longer axes of neighboring elliptical waveguides without reducing the number of guiding elements. This allows us to study which vortex solitons are allowed in such structures and how exactly the splitting of higher-order singularities occurs when the state is forbidden by the discrete rotational symmetry of the array. In addition, we show that the number of stable vortices substantially decreases with decrease of the order of discrete rotation symmetry of such array.

We consider paraxial propagation of a light beam along the $z$-axis of focusing cubic nonlinear medium with imprinted transverse refractive index modulation. The evolution of the scaled amplitude of the light field $\psi$ can be described by the following nonlinear Schrödinger equation:

$$i\frac{\partial \psi}{\partial z} = -\frac{1}{2}\left(\frac{\partial^2 \psi}{\partial x^2} + \frac{\partial^2 \psi}{\partial y^2}\right) - \mathcal{R}(x,y)\psi - |\psi|^2\psi, \quad (1)$$

where the transverse coordinates $x,y$ are normalized to the characteristic scale $r_0 = 10\ \mu\text{m}$, the propagation distance $z$ is normalized to the diffraction length $kr_0^2 \approx 1.44\ \text{mm}$ (here we assume working wavelength of $\lambda = 800\ \text{nm}$), $k = 2\pi n/\lambda$ is the wavenumber in the medium, $n \approx 1.45$ is the unperturbed refractive index. Here we aim to create the structure, whose discrete rotational symmetry can be adjusted without changing the number of the guiding channels in the system. This is achieved by composing ring arrays $\mathcal{R}(x,y)$ from identical, but mutually rotated elliptical waveguides $\mathcal{Q}(x,y) = pe^{-(x/\sigma_1)^2-(y/\sigma_s)^2}$ with widths $\sigma_1 = 0.70$ and $\sigma_s = 0.35$ along the long and short axes, respectively, and equal depth $p = k^2r_0^2\delta n/n$ defined by the refractive index modulation depth $\delta n$. Such elliptical waveguides naturally form when one uses direct fs-laser writing technique, while the orientation of waveguide's longer and shorter axes can be controlled by changing the orientation of the sample with respect to writing beam, see review [6]. Further we consider the waveguides with $p = 5$ that corresponds to $\delta n \approx 5.6 \times 10^{-4}$. We employ $N$ such waveguides $\mathcal{Q}(x,y)$ placed equidistantly on a ring of radius $r = N/2$, but we suppose that longer axis of each subsequent waveguide with index $k = 0, ..., N-1$ (counted in the counterclockwise direction starting from the right outermost waveguide corresponding to $k = 0$, which is always oriented parallel to the $x$-axis) is twisted with respect to longer axis of previous waveguide by the angle $\phi_r$ [see Fig. 1(a)-1(e), top row, for examples of structures corresponding to different specially selected twist angles $\phi_r$].

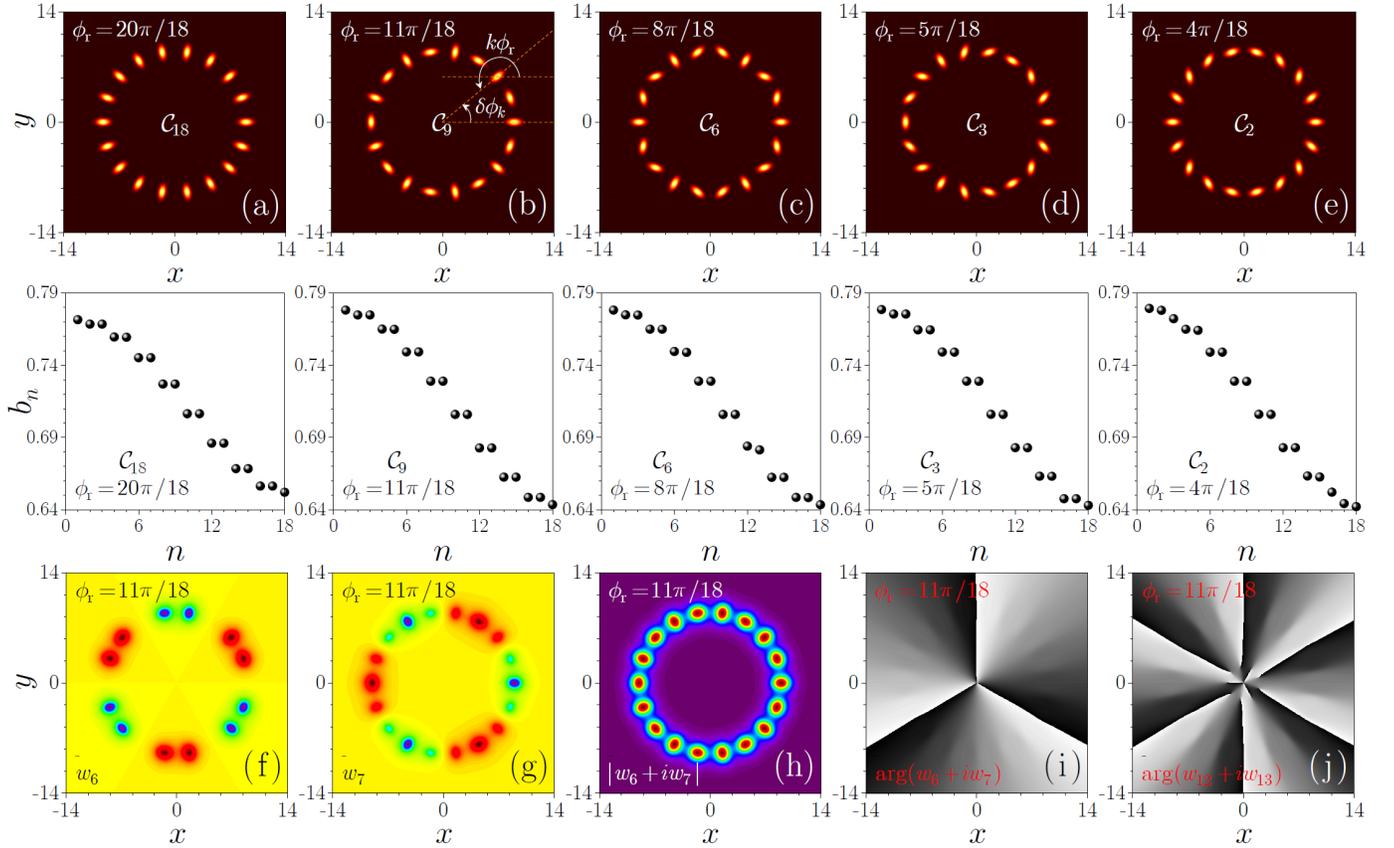

Fig. 1. (a)-(e) Examples of circular waveguide arrays $\mathcal{R}$ (top row) composed from $N = 18$ elliptical waveguides for different twist angles $\phi_r$ between neighboring waveguides indicated on the panels, and representative spectra of linear modes (middle row) supported by such arrays. The order of discrete rotational symmetry is indicated for each array. (f),(g) Representative degenerate eigenmodes $w_{n=6,7}$ supported by $\mathcal{C}_9$ array, and field modulus (h) and phase (i) distributions in $m = 3$ vortex produced by linear combination $w_6 + iw_7$ of these modes. (j) Phase structure of the field produced by linear combination $w_{12} + iw_{13}$. Here and below $p = 5$, and widths of waveguides along the long and short axes are $\sigma_1 = 0.70$, $\sigma_s = 0.35$.

Thus, the total twist angle with respect to $x$-axis for the waveguide with index $k$ is equal to $k\phi_r$, as indicated in Fig. 1(b). At the same time, the angle between the $x$-axis and the radial line passing through the center of the waveguide with index $k$ is given by $\delta\phi_k = 2\pi k/N$, because waveguides are placed equidistantly on the ring, see Fig. 1(b). One can see, that the array $\mathcal{R}(x,y)$ possesses discrete angular rotation symmetry only for a limited set of twist angles, when (i) the condition $k\phi_r - \delta\phi_k = \pi$ is satisfied (in this case, longer axis of the waveguide with index $k$ points toward the center of the ring, and the same happens for longer axes of the waveguides with indices $2k$, $3k$, etc.), and (ii) when total number of waveguides $N$ is simultaneously divisible by this $k$ value. This yields the expression for the twist angle:

$$\phi_{\rm r} = \frac{\pi(N+2k)}{kN}, \qquad (2)$$

while discrete rotational symmetry of the array satisfying two above conditions is given by $\mathcal{C}_{N/k}$. Thus, by varying the twist angle for elliptical waveguides, one can control discrete rotational symmetry of the array without changing the number $N$ of guiding elements in it. The examples of such arrays with discrete rotational symmetries ranging from $\mathcal{C}_{18}$ ($k=1$) to $\mathcal{C}_2$ ($k=9$) are presented in the top row of Fig. 1(a)-1(e) for sufficiently large $N=18$ (the order of discrete rotational symmetry and twist angles are indicated for each structure).

To study how discrete rotational symmetry affects linear spectra of such arrays, we search for their linear eigenmodes $\psi = w(x,y)e^{ibz}$, where $w(x,y)$ is the real function describing mode profile, and $b$ is its propagation constant. Substituting the field in such form into linearized version of Eq. (1) (with omitted nonlinear term), we obtain the linear eigenvalue problem that was solved numerically using plane-wave expansion method. We used up to 300*300 spatial harmonics to ensure accurate calculation of the spectrum. Because waveguides are single-mode, the discrete spectrum consists of $N$ eigenvalues, where largest eigenvalue $b_1$ corresponds to the mode, where field in all waveguides is in-phase, while lowest eigenvalue $b_N$ corresponds to the mode, where field in neighboring waveguides is out-of-phase (for even $N$). In structure with largest discrete rotational symmetry $\mathcal{C}_{18}$ ($k=1$) all eigenvalues except for the first and last ones form degenerate pairs [see linear spectrum in Fig. 1(a), middle row]. Combinations of modes from degenerate pairs allow to construct vortices, where, for example, the combination $w_2 + iw_3$ produces vortex with topological charge 1, while combination $w_{16} + iw_{17}$ produces vortex with charge 8 with singularity in the center. As mentioned above, next mode $w_{18}$ is a multipole mode that does not carry vorticity. This means that maximal available charge of the vortex states in such arrays is limited, in full agreement with the results of [42,43], obtained using group theory analysis, where it was shown that in arrays with discrete rotational symmetry of the order $\mathcal{C}_N$ only vortices with topological charges $|m| < N/2$ can exist, for even $N$, while for odd $N$ one has $|m| < (N+1)/2$.

When the order of discrete rotational symmetry in our array decreases for smaller twist angles $\phi_{\rm r}$ of elliptical waveguides (i.e. for larger $k$ values), the total number of linear modes supported by the ring array remains unchanged, but the degeneracy of some eigenvalues can be lifted. This is particularly well visible for eigenvalues $b_{12}$ and $b_{13}$ in structure with discrete rotational symmetry $\mathcal{C}_6$ [Fig. 1(c), middle row], or for eigenvalues $b_2, b_3$ and $b_{16}, b_{17}$ in structure with discrete rotational symmetry $\mathcal{C}_2$ [Fig. 1(e), middle row]. In general, the eigenvalues in pairs, located in the top and bottom parts of the spectrum, that were degenerate in $\mathcal{C}_{18}$, experience largest shifts/splitting in structures with low discrete rotational symmetry, like $\mathcal{C}_2$, while in the depth of the spectrum the splitting of eigenvalues is usually not so pronounced and sometimes may be rather small, at the level of numerical accuracy of eigenvalue calculations (sometimes the difference appears in sixths-eights decimal places). The modes, whose linear combination produces vortices *allowed* for a given discrete rotational symmetry of the structure remain degenerate. The examples of two such modes $w_6$ and $w_7$ in $\mathcal{C}_9$ structure are presented in Fig. 1(f) and 1(g), while field modulus and phase distributions for charge-3 vortex, obtained as linear combination $w_6 + iw_7$, are presented in Fig. 1(h) and 1(i), respectively. Notice that the distribution $|w_6 + iw_7|$ inherits discrete rotational symmetry of the array. A somewhat surprising result is that pairs of eigenvalues for modes, whose combinations should correspond to vortices with *forbidden* topological charges, may also remain degenerate upon reduction of rotational symmetry. This means that symmetry reduction for the array does not necessarily leads to lifting of degeneracy for *all* modes that would produce forbidden vortices. Instead, combinations of such modes yield vortical structures with *split* singularities, see example in Fig. 1(j), where phase distribution is shown for combination $w_{12} + iw_{13}$ in $\mathcal{C}_9$ array (the same combination in $\mathcal{C}_{18}$ array produces single charge-6 vortex in the center). Split singularities shift away from the center of the ring. If one calculates phase accumulation on the ring of sufficiently large radius, one still gets $2\pi m$, even though $m$ does not determine now the charge of central singularity, but instead characterizes global vorticity of solution – on this reason we further call such states as $\psi_m$ states.

It should be stressed that even if longer axis of the right outermost waveguide with index $k=0$ is not parallel to the $x$ axis, the corresponding arrays constructed for twist angles (2) still feature the same discrete rotational symmetry as their counterparts from Fig. 1, where $k=0$ waveguides are parallel to the $x$ axis. Linear spectra of modes supported by former arrays are practically identical to spectra shown in the middle row of Fig. 1 (except for small vertical shift) and, consequently, splitting of singularities in vortex solitons with forbidden charges in arrays with different orientations of waveguide's longer axes with respect to the ring are nearly the same.

Next, we study how splitting of singularities obvious even in linear case, affects properties of available vortex solitons in such arrays. We search for profiles of vortex solitons supported by arrays with different discrete rotational symmetry in the form $\psi = (w_{\rm re} + iw_{\rm im})e^{ibz}$, where the functions $w_{\rm re,im}(x,y)$ describe real and imaginary parts of solution, while $b$ is the nonlinear propagation constant. This substitution into Eq. (1) yields the system of equations:

$$bw_{\rm re} = \frac{1}{2}\left(\frac{\partial^2 w_{\rm re}}{\partial x^2} + \frac{\partial^2 w_{\rm re}}{\partial y^2}\right) + \mathcal{R}w_{\rm re} + (w_{\rm re}^2 + w_{\rm im}^2)w_{\rm re},$$
$$bw_{\rm im} = \frac{1}{2}\left(\frac{\partial^2 w_{\rm im}}{\partial x^2} + \frac{\partial^2 w_{\rm im}}{\partial y^2}\right) + \mathcal{R}w_{\rm im} + (w_{\rm re}^2 + w_{\rm im}^2)w_{\rm im}, \qquad (3)$$

that was solved using Newton method. In this method we used standard 5-point discretization scheme for spatial derivatives. The tolerance used upon calculation of soliton solutions was as low as $10^{-12}$. The obtained soliton families are characterized by the dependence of power $U$ and $z$-component of the orbital angular momentum $L$:

$$U = \int_{-\infty}^{\infty} dy \int_{-\infty}^{\infty} |\psi|^2 dx,$$
$$L = \mathbf{e}_z \mathbf{L} = \frac{\mathbf{e}_z}{2i} \int_{-\infty}^{\infty} dy \int_{-\infty}^{\infty} [\mathbf{r} \times (\psi^* \nabla \psi - \psi \nabla \psi^*)] dx, \qquad (4)$$

where $\nabla = (\partial/\partial x, \partial/\partial y)$, on propagation constant $b$.

In Fig. 2 we first show the families of vortex solitons in the array with largest discrete rotational symmetry $\mathcal{C}_{18}$, whose order coincides with the number of guiding channels. Such solitons bifurcate from linear guided vortex modes of the structure, so power of corresponding solitons $\psi_m$ vanishes when propagation constant $b$ approaches corresponding eigenvalue (cutoff), see Fig. 2(a). The higher is the charge $m$ of soliton, the lower its cutoff propagation constant. Therefore, for fixed $b$ vortex solitons with higher topological charges have higher power. The examples of field modulus and phase distributions in vortex solitons in $\mathcal{C}_{18}$ array are presented in Fig. 2(c). Notice that field modulus distributions of all such states inherit discrete rotation symmetry of the array, as highlighted by dashed while lines for $\psi_1$ state. The depth of azimuthal modulation in $|\psi_m|$ distributions increases with increase of the topological charge $m$. Since in this array all vortices with charges up to $m=8$ are allowed by symmetry of the structure, in each case one observes one higher-order phase singularity in the center of ring array (small splitting visible for highest charges is due to finite number of transverse points used in simulations, in our case $1001 \times 1001$, and it gradually vanishes with further increase of the number of points). Even though orbital angular momentum $L$ is not a conserved quantity in the presence of the external optical potential $\mathcal{R}$ (for arbitrary inputs), its dependence on $b$ shown in Fig. 2(b) indicates that in the array vortex solitons with higher changes may, in principle, carry lower momentum for fixed $b$, in contrast to situation encountered in uniform cubic medium, where the relation $L = mU$ holds and higher-charge states carry larger momentum at fixed $b$.

We are also interested in the influence of discrete rotational symmetry of stability properties of vortex solitons. Stability was tested by adding small broadband perturbations into input field distributions and propagating them in the frames of Eq. (1) up to $z = 10^4$ that reveals the presence of even weak instabilities, if solution is unstable. To model soliton propagation, we used standard split-step fast Fourier transform method with longitudinal step $dz = 10^{-3}$. In Fig. 2(a) and 2(b) black lines mark stable families, while red lines mark unstable families. One can see that only vortex solitons with highest topological charges are stable (in this case, with $m > 4$), while low-charge vortices with $m \leq 4$ are unstable.

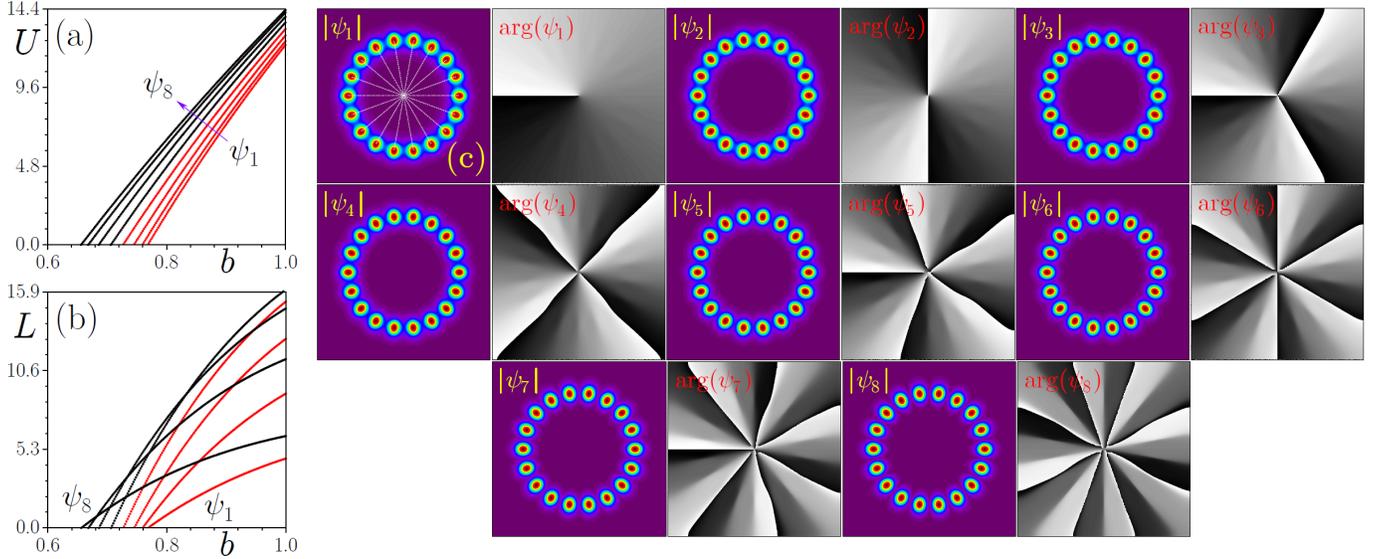

Fig. 2. (a) Power and (b) angular momentum of different vortex-carrying modes versus propagation constant in waveguide array with $N = 18$ elliptical waveguides, with twist angle $\phi_r = 20\pi/18$ ($\mathcal{C}_{18}$ discrete rotational symmetry). Stable families are shown black, unstable families are shown red. (c) Examples of field modulus (left) and phase (right) distributions in vortex-carrying soliton solutions from $\psi_1$ to $\psi_8$ with $b = 0.9$ shown within the window $x, y \in [-14, +14]$. White dashed lines in panel for $\psi_1$ solution illustrate discrete rotational symmetry of field modulus distribution.

Next, we study how the above picture changes when the order of discrete rotational symmetry of the structure decreases. In Fig. 3 we summarize the properties of vortex solitons supported by the array with lower, $\mathcal{C}_9$, symmetry achieved for $\phi_r = 11\pi/18$. In this case, charge rule proposed in [42,43] suggests that solitons with single high-charge singularity in the center of the array can exist only for $|m| < 5$. Indeed, phase distributions of the available vortex solitons in this array shown in Fig. 3(c) reveal instead of single central singularity the presence of several charge-1 singularities in "forbidden" states, from $\psi_5$ to $\psi_8$, i.e. central singularity in them experiences splitting. Nevertheless, they still exist as self-sustained states, but with more complex phase distributions.

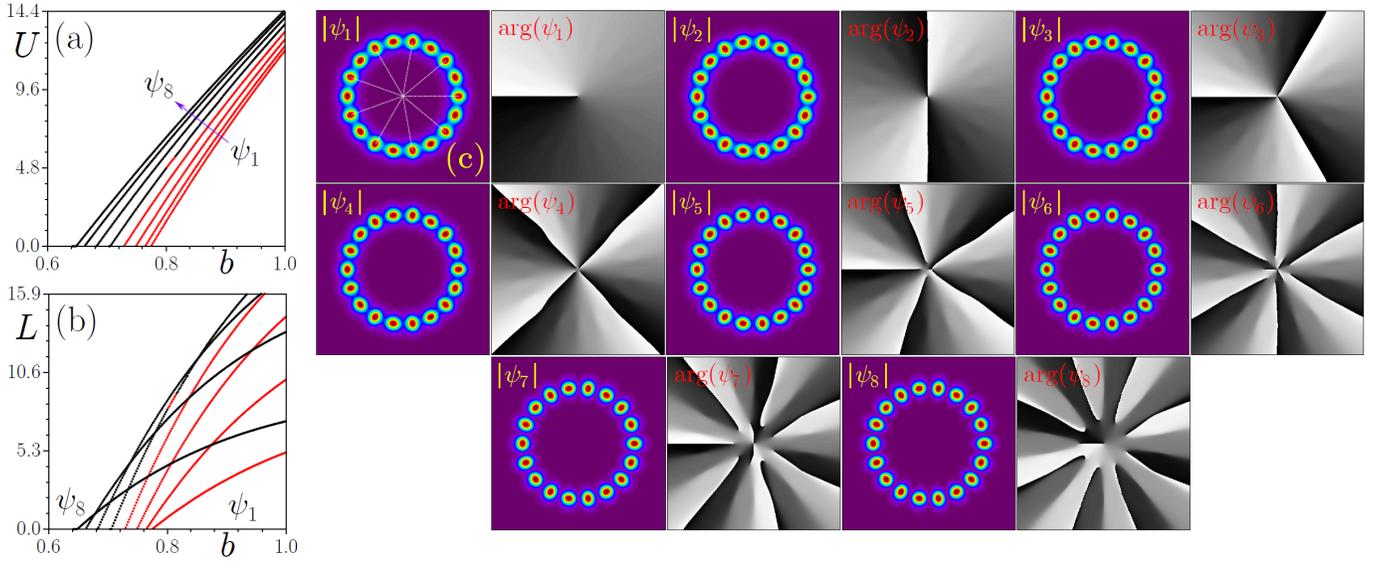

Fig. 3. The same as in Fig. 2, but for waveguide array with twist angle $\phi_r = 11\pi/18$ featuring $\mathcal{C}_9$ discrete rotational symmetry. Notice progressively increasing splitting of vortices starting from $\psi_5$ state.

The splitting (i.e. the distance between charge-1 singularities and center of the ring) of singularities in latter states progressively increases with the order of solution $\psi_m$ (for $m \geq 5$). Additional singularities may appear also near the center of the array but global vorticity defined over the ring of sufficiently large radius still equals to $m$. The dependencies $U(b)$ and $L(b)$ remain qualitatively similar in this case [Fig. 3(a) and 3(b)], but one can see the emergence of small instability segment for $\psi_5$ state.

Further reduction of discrete rotational symmetry down to $\mathcal{C}_6$ for $\phi_r = 8\pi/18$ results in much larger splitting of singularities in $\psi_5$ soliton and the onset of splitting in $\psi_4$ soliton, see Fig. 4(a) and 4(b) for soliton families and Fig. 4(c) for examples of field modulus and phase distributions. Surprisingly, even though charge rule predicts that in this case solitons with single high-order phase singularity can exist only at $|m| < 3$, we do not observe noticeable splitting of phase singularity in $\psi_3$ state (we attribute this to sufficiently large radius of ring array considered here and to the fact that pronounced splitting of singularities with low charges requires much stronger azimuthal modulations of the structure than those considered here). At the same time, the reduction of discrete rotational symmetry strongly affects stability of solutions, since now the branch of $\psi_5$ solitons becomes unstable in the entire interval of propagation constants shown in Fig. 4(a).

The study of other arrays with even lower discrete rotational symmetries down to $\mathcal{C}_2$ confirms the tendency for more pronounced splitting of singularities and destabilization of $\psi_m$ solitons of higher orders $m$. Thus, in $\mathcal{C}_3$ array, realized for $\phi_r = 5\pi/18$, even the $\psi_6$ branch acquires large instability segments (the branches from $\psi_1$ to $\psi_5$ are unstable too), while $\psi_7$ branch features only small instability segment near the cutoff.

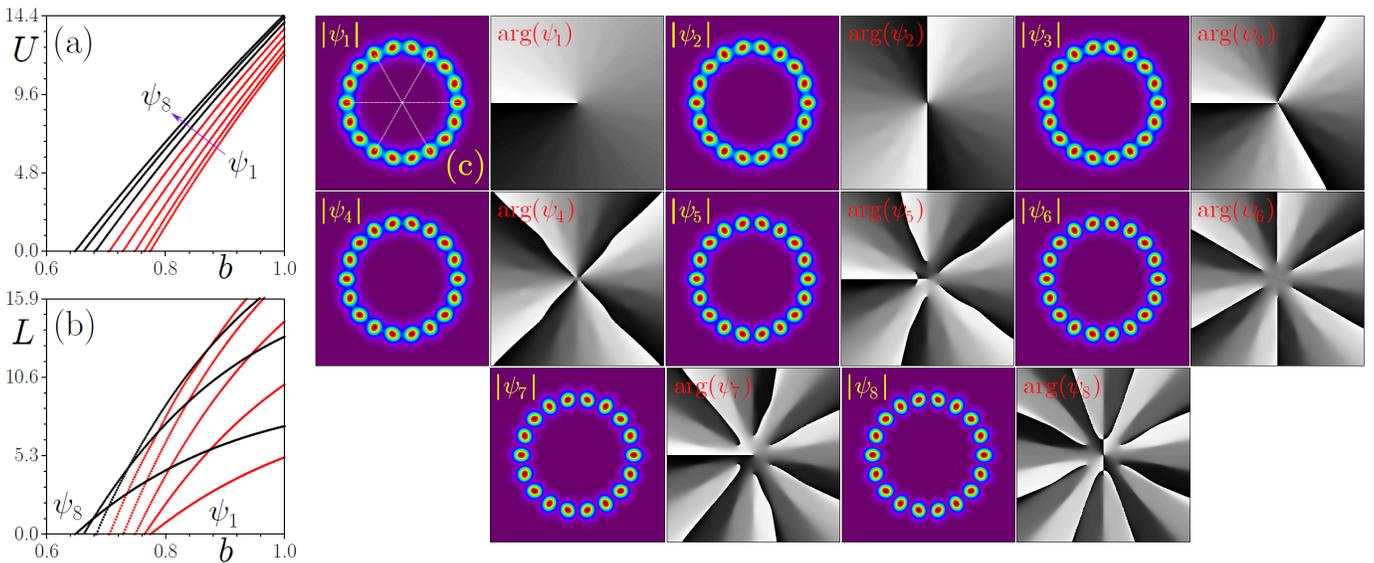

Fig. 4. The same as in Fig. 2, but for waveguide array with twist angle $\phi_r = 8\pi/18$ featuring $\mathcal{C}_6$ discrete rotational symmetry.

Typical scenarios of stable and unstable propagation of perturbed vortex solitons are shown in Fig. 5, on the example of $\mathcal{C}_6$ array. Thus, upon development of the instability of $\psi_3$ and $\psi_5$ solitons, illustrated in Fig. 5(a) and 5(b), their peak amplitude $a = \max|\psi|$ first gradually increases and then starts performing irregular oscillations. This is usually accompanied by loss of vortical phase structure and appearance of strong azimuthal intensity modulations running along the ring of waveguides. In contrast, stable vortex solitons, even when perturbed, show only small-amplitude oscillations and maintain their internal structure up to large propagation distances, see example for $\psi_6$ soliton in Fig. 5(c).

Summarizing, we have studied the impact of discrete rotational symmetry of the system that can be adjusted in large-scale ring array proposed here without changing the number of guiding elements in it, on the structure and stability properties of vortex solitons. We have found a general tendency for splitting of high-order phase singularities into sets of charge-1 off-center singularities when the former are forbidden by discrete rotational symmetry of the array. The splitting of singularities increases with increase of the order (or global vorticity) of the soliton solution. At the same time, our results demonstrate the existence of a rich variety of stable vortex solitons with complex internal phase structure. Stability analysis shows that decrease of the order of discrete rotational symmetry usually leads to destabilization of the families of vortex solitons with lowest charges, while the states with highest global vorticity remain stable. It is relevant to mention that our results can be potentially extended to subwavelength systems, such as arrays of elliptical coupled metallic nanowires, where formation of discrete fundamental and vortex solitons may be possible, as shown in [60,61]. Specific internal structure of eigenmodes of individual nanowires in such system may dramatically enrich the families of vortex solitons bifurcating from linear modes of circular arrays.

**Acknowledgements:** This research is funded by the research project FFUU-2021-0003 of the Institute of Spectroscopy of the Russian Academy of Sciences.

**Data availability**
Data will be made available upon reasonable request.

**Declaration of Competing Interest**
The author declares that there is no conflict of interests.

**Credit authorship contribution statement**
The manuscript was written and all results were obtained by the only author.

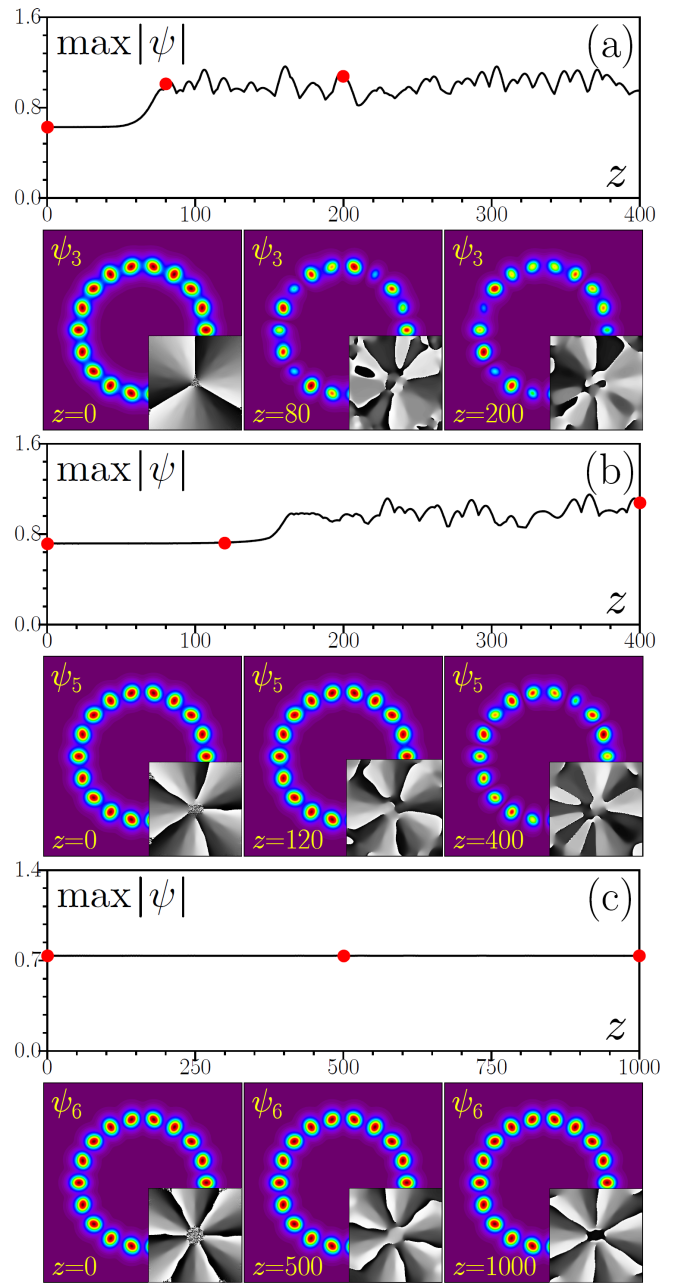

Fig. 5. Propagation dynamics of perturbed unstable $\psi_3$ (a) and $\psi_5$ (b) solitons, and of stable $\psi_6$ (c) soliton with $b = 0.9$ in waveguide array with twist angle $\phi_r = 8\pi/18$ featuring $\mathcal{C}_6$ discrete rotational symmetry. In each case top rows show peak amplitude versus distance $z$, while bottom rows show field modulus and phase (insets) distributions within $x, y \in [-14, +14]$ window corresponding to the red dots in $\max|\psi(z)|$ dependence.